\documentstyle[]{article}

\begin{document}
\begin{center} {\Large {\bf Actio causes reactio: Gravito-optical
trapping\\ of three-level atoms\footnote{to appear in Phys. Rev. A}
}}\\[1cm]
Karl-Peter Marzlin
\footnote{e-mail: peter.marzlin@uni-konstanz.de}
and J\"urgen Audretsch
\footnote{e-mail: juergen.audretsch@uni-konstanz.de}
\\[2mm]
Fakult\"at f\"ur Physik
der Universit\"at Konstanz\\
Postfach 5560 M 674\\
D-78434 Konstanz, Germany
\end{center}
$ $\\[3mm]
\begin{minipage}{15cm}
\begin{abstract}
We investigate an atomic three-level $\Lambda$-system which
is exposed to
two counterpropagating laser fields (inducing Raman
transitions) and which is closed by
a magnetic hyperfine field tuned to be
in resonance with the transition between the two ground states.
The influence of a homogeneous gravitational field is included
in a full quantum treatment of the internal and external dynamics
of the atom.
It is shown that the combined influence of the gravitational field
and the lasers lead for specific momentum values with a very high
probability to a transition of the Landau-Zener type. This is
accompanied by a momentum transfer resulting in an upward kick.
For appropriate initial conditions a sequence of up and down motions
is obtained. No mirror is needed.
A gravito-optical trapping of atoms based on this effect seems to
be realizable.
\end{abstract}
\end{minipage}
$ $ \\[1cm]
\section{Introduction}
During the last decade new techniques in atom optics led to the
possibility to produce atomic clouds and beams with very small
velocity (see Ref. \cite{mlynek94} and references therein).
Under such circumstances the influence of gravity has to be taken
into account. To see in detail how gravity changes the atomic
center-of-mass motion in the presence of laser fields,
atoms exposed simultaneously
to a running laser wave and gravity were theoretically
studied in Refs. \cite{marzlin96,labo95}.
In many experiments gravity causes an unwanted
change in the atomic momentum that must be accounted for in the
design of an experiment (see, e.g., Ref. \cite{balykin89}).
But there are several proposals and realizations in which
gravity is used to slow or cool atoms. One example is the
gravitational cavity \cite{aminoff93,wallis92,liston95} in which
freely falling atoms are reflected by an atomic mirror at the
bottom of the cavity. Here gravity is directly used to reverse the
atomic upward motion. Another example is the gravitational Sisyphus
cooling \cite{newbury95,pritchard93}
in which a magnetic field gradient
used to cancel gravity produces a cooling force that is proportional
to gravity. Another proposal \cite{klimov95} exploits the fact
that the addition of the gravitational potential to a gradient
force potential can result in new minima which may be used to
form a trap.

The gravitational cavity designed to trap atoms which is
discussed in the
references given above is based on the multiple bouncing of atoms
on a reflective surface. It essentially contains two elements: a
lower mirror for atoms provided by an evanescent laser wave and
the influence of gravity which results in a second upper "mirror"
closing the cavity because it bends the atomic trajectories. By
this a "trampoline" geometry is obtained. In the following we want
to present a
setup in which the lower mirror closing the trap at the bottom is
replaced by a combined influence of the gravitational field and
the laser
fields. Gravitation as "{\em actio}" causes a free fall
which together with the influence of the lasers leads to a
"{\em reactio}" of the atoms turning them into an upward motion.
A gravito-optically induced internal transition
of the atoms is accompanied by a momentum transfer which causes an
upward kick. Afterwards, the free fall of the atoms continues until
at a certain
downward momentum resonance is obtained again, the next
kick happens, and so on. In this way a
sequence of up and down motions is obtained without any
reflection at a surface. The result obtained could
be called "trampolining without trampoline".

In detail we study the free fall of a three-level $\Lambda$-system
interacting with two laser fields and a magnetic field which is
applied between the two lower levels, compare Fig. 1. In section 2
we perform the rotating-wave-approximation and switch to the
interaction picture. In section 3 the excited state is adiabatically
eliminated. It turns out that in momentum space the dynamical
equation separates into differential equations coupling ladders
of states. It becomes evident that the fundamental gravito-optical
transitions are of Landau-Zener type. The resulting
center-of-mass motion of the atoms is read off in section 4 and
a simple visualization of the underlying physics is given. A
complete description of the process with reference to dressed states
(interaction energy of the magnetic field included) is added in
section 5. We point out in section 6 that the setup represents a
trap for atoms. If the initial state is a momentum state, drops fall
out of the trap. If this assumption is not made, the trap is
continuously leaking atoms.
\section{Transformed Hamiltonian}
As seen from the inertial system the dynamical evolution
of the three-level system in question
is governed by the Hamiltonian
\begin{equation}
H := H_A + H_{c.m.} + H_{int}
\label{ham} \end{equation}
where
\begin{equation}
H_A := E_e | e \rangle \langle e | + E_+ | + \rangle \langle +|
      + E_- |- \rangle \langle -|
\end{equation}
describes the internal energy levels $E_e > E_+ > E_- $ of the
atom,
\begin{equation}
H_{c.m.} := {\bf 1} \Big \{ \frac{\vec{p}^2}{2M} - M \vec{a}\cdot
 \vec{x} \Big \}
\end{equation}
is the center-of-mass part of the complete Hamiltonian, and
\begin{eqnarray}
H_{int} &:=& -\hbar \Omega \Big \{ \cos [\omega_+ t - \vec{k}_+
    \cdot \vec{x} -\varphi_+] (| e\rangle\langle +| \, +\,
    |+ \rangle \langle e| )  \nonumber \\ && \hspace{1cm} +
    \cos [\omega_- t - \vec{k}_-
    \cdot \vec{x} -\varphi_-] (| e\rangle\langle -|\,  +\,  |-\rangle
    \langle e | ) \Big \}  \nonumber \\ && +
    \hbar \Omega_B  \cos [\omega_B t - \varphi_B] (|+\rangle
    \langle -|\,  +\,  |-\rangle\langle +|)
\end{eqnarray}
describes the influence of the two laser fields and the magnetic
hyperfine field on the atom. $\vec{x}$ and $\vec{p}$ are the
position and momentum operator of the atom's center-of-mass motion
and $M$ is its mass. The homogeneous
gravitational acceleration is denoted by the vector
$\vec{a}$ pointing towards the Earth. $\omega_{\pm}$
are the frequencies of the two Raman lasers which induce
transitions between the upper state $| e\rangle$ and the lower
states $|+\rangle$
and $|-\rangle$, respectively, see Fig. 1.
Their phases are given by $\varphi
_{\pm}$, and $\vec{k}_{\pm}$ are their counterpropagating
wave vectors. The Rabi
frequency $\Omega$ is assumed to be equal for both laser fields.
$\omega_B$, $\Omega_B$, and $\varphi_B$ denote the frequency, the
Rabi frequency, and the phase of the magnetic field.
${\bf 1}$ is the unit operator in the internal three dimensional
space. No
spontaneous emission is included in $H$ since we will work in a
regime where it can be neglected.

By a sequence of unitary transformations we will
now transform $H$ into
a form which will allow us to read off the physical content
more easily. First we
perform the rotating wave approximation by using
\begin{equation}
U_1 := |e\rangle\langle e|\,  +\,  |+\rangle\langle +|
   e^{i \omega_+ t} + \, |-\rangle\langle - | e^{i\omega_- t}
\end{equation}
to transform the original state vector $| \psi \rangle$ to
$| \psi_1 \rangle = U_1^+ |\psi \rangle$. Throughout the paper
we will use the convention that a unitary transformation
$U$ acts always in the form $|\psi_{old} \rangle = U |\psi_{new}
\rangle$. Neglecting terms oscillating with frequency
$2 \omega_{\pm}$ and $2 \omega_B$ we obtain
\begin{eqnarray}
H_1 &=& E_e {\bf 1} + \hbar \Delta_+ |+ \rangle\langle +| \, +
  \hbar \Delta_- |-\rangle\langle -|\,  + H_{c.m.} \nonumber\\ &&
  -\frac{\hbar\Omega}{2} \Big \{ e^{i(\vec{k}\cdot\vec{x} +
   \varphi_+)} |e\rangle\langle +| \, + e^{i(-\vec{k}\cdot\vec{x} +
   \varphi_-)} |e\rangle\langle -| \, + H.c. \Big \} \nonumber\\ &&
   +\frac{\hbar\Omega_B}{2} \Big \{ e^{i \varphi_B}
  |+\rangle\langle -| \,  + H.c. \Big \}
\label{ham1}\end{eqnarray}
Here we have set $\vec{k} := \vec{k}_+$ and have made use of the
fact that the laser beams are counterpropagating and have about
the same frequency so that $\vec{k}_-$ is approximately equal to
$- \vec{k}$.

Note that we do {\em not} set $\omega_+ = \omega_-$ in the
calculations. This ambiguous treatment of frequency and wavelength
of the lasers deserves a short comment. Setting $\vec{k}_+ =
-\vec{k}_-$ is of course an approximation which affects the
momentum conservation. But since the corresponding
error $\delta \vec{k} := \vec{k}_+ + \vec{k}_-$
is very small compared to the vectors themselves
this approximation is justified as far as momentum conservation
is concerned and as long as we do not look at momenta which are of
the order of $ \hbar \delta \vec{k}$.
But setting $\omega_+ = \omega_-$ would violate the energy
conservation by an amount of about $E_+-E_-$. Since also $\hbar
\omega_B$ and many other energy scales occurring in the system at
hand lie in this range, this approximation would be unacceptable.
The error
\begin{equation} \frac{\hbar^2 (\vec{k} + \delta \vec{k})^2}{2M}-
  \frac{\hbar^2\vec{k}^2}{2M} \approx \frac{1}{M}\hbar^2 \vec{k}\cdot
  \delta \vec{k} = \frac{\hbar \omega_+ (\hbar \omega_+-\hbar
  \omega_-)}{Mc^2} \end{equation}
in the kinetic energy caused by $\delta \vec{k}$ is negligible,
however.

To achieve the time independence of $H_1$ it was
necessary to impose the condition
\begin{equation}
\omega_B + \omega_+ - \omega_- = 0
\label{freqbed} \end{equation}
on the field frequencies. This also allows us to perform the
rotating wave approximation simultaneously in $\omega_{\pm}$
and $\omega_B$. In the following we will restrict to setups where
this is fulfilled.
To facilitate the calculations we furthermore assume that the
detunings $\Delta_{\pm} := \omega_{\pm} - (E_e -E_{\pm})/\hbar$
of the two laser beams are equal: $\Delta_+ = \Delta_- =: \Delta$.
Eq. (\ref{freqbed}) then implies that the magnetic field has to
be in resonance with the hyperfine transition,
\begin{equation}
\omega_B = \frac{E_+ - E_-}{\hbar}  \; .
\label{obres} \end{equation}

It proves to be useful to perform a second unitary transformation
with the operator
\begin{equation}
U_2 = e^{-itE_e/\hbar} \Big \{
  |e\rangle\langle e|\,  +\,  |+\rangle\langle +|
   e^{-i \varphi_+} + \, |-\rangle\langle - | e^{-i\varphi_-}
   \Big \} \exp [i M \vec{a}\cdot \vec{x}t/\hbar] \; .
\end{equation}
The first factor shifts the overall energy of the internal states,
the second term removes the phase factors from the Raman transition
matrix elements, and with the last term we switch to the interaction
picture with respect to the gravitational potential. This leads to
\begin{eqnarray}
H_2 &=& {\bf 1} \frac{(\vec{p}+ M \vec{a}t)^2}{2M} + \hbar \Delta
   \Big \{ |+\rangle\langle +| \, + \,
   |-\rangle\langle -|\Big \} \nonumber \\ &&
   - \frac{\hbar \Omega}{2} \Big \{ e^{i \vec{k}\cdot \vec{x}}
   |e\rangle\langle +| \, + e^{-i \vec{k}\cdot \vec{x}}
   |e \rangle\langle -| \, + H.c. \Big \}
   +\frac{\hbar \Omega_B}{2} \Big \{ e^{i \Delta \varphi}
  |+\rangle\langle -| \, + H.c. \Big \}
\label{h2} \end{eqnarray}
with $\Delta \varphi := \varphi_B +\varphi_+ - \varphi_-$.
\section{Reduction of the Schr\"odinger equation}
To solve the Schr\"odinger equation it is advantageous to expand
the wave function in momentum space according to
\begin{equation}
|\psi_2\rangle = \int d^3 q | \vec{q}\rangle \otimes \Big \{
  c_e(\vec{q}) e^{i \phi(t)}|e\rangle + c_+(\vec{q}) e^{i(\phi(t)
  -\Delta t)} |+\rangle + c_-(\vec{q})e^{i(\phi(t)-\Delta t)}
  |-\rangle \Big \} \label{psi2} \end{equation}
with
\begin{equation}
  \phi(t) := -\frac{1}{\hbar}\Big\{\frac{\vec{q}^2}{2M}t+
  {1\over 2} \vec{q}\cdot \vec{a} t^2
  + \frac{M}{6} \vec{a}^2 t^3 \Big \} \end{equation}
where $| \vec{q} \rangle$ are the eigenvectors of the untransformed
momentum operator $\vec{p}$ with
$\vec{p}| \vec{q} \rangle = \vec{q} | \vec{q} \rangle$,
and the $c(\vec{q})$ are functions of
the time $t$. Remember that we already have made two unitary
transformations. The measured mean value of the momentum of the
states $ |\vec{q} \rangle$ is therefore
$\langle \vec{q} | \vec{p}_2 | \vec{q} \rangle = \vec{q} + M
\vec{a} t$. Consequently the center -of-mass part $ |\vec{q} \rangle$
of $|\psi_2 \rangle$
describes states already "falling" under the influence of gravity
according to the classical law of free fall which in this way has
been separated in the subsequent calculation. $\vec{q}$ is a
momentum
parameter which agrees with the measured momentum at $t=0$.

The Schr\"odinger equation is then reduced to
\begin{eqnarray}
  i \dot{c}_e(\vec{q}) &=& \frac{-\Omega}{2} e^{-i\Delta t}
  e^{-i \delta_R t} \Big \{ c_+(\vec{q}- \hbar \vec{k})
  e^{i(\vec{q}\cdot\vec{k}t/M + \vec{k}
  \cdot \vec{a}t^2/2)} + c_-(\vec{q}+\hbar \vec{k})
  e^{-i(\vec{q}\cdot\vec{k}t/M
  +\vec{k} \cdot \vec{a}t^2/2)}\Big \}
  \label{bewgl1} \\
  i\dot{c}_+(\vec{q}) &=& \frac{-\Omega}{2} c_e(\vec{q}+\hbar
  \vec{k}) e^{i\Delta t}
  e^{-i(\vec{q}\cdot\vec{k}t/M + \delta_R t + \vec{k}
  \cdot \vec{a}t^2/2)} + \frac{\Omega_B}{2} e^{i\Delta\varphi}
  c_-(\vec{q}) \label{bewgl2}\\
  i\dot{c}_-(\vec{q}) &=& \frac{-\Omega}{2} c_e(\vec{q}-\hbar
  \vec{k}) e^{i\Delta t}
   e^{i(\vec{q}\cdot\vec{k}t/M - \delta_R t + \vec{k}
  \cdot \vec{a}t^2/2)} + \frac{\Omega_B}{2} e^{-i\Delta\varphi}
  c_+(\vec{q}) \label{bewgl3}
\end{eqnarray}
$\delta_R := \hbar \vec{k}^2/(2M)$ is the recoil shift of the atom.

This form is particularly suited for the adiabatic elimination
of the excited state. We obtain an effective two-level system
with internal states $|+ \rangle$ and $|- \rangle$.
This can be performed
if the detuning $\Delta$
is much larger than all other frequencies appearing in
Eqs. (\ref{bewgl1}) to (\ref{bewgl3}).
Proceeding as in Ref. \cite{moler92} we
assume that the time dependence of the r.h.s. of Eq. (\ref{bewgl1})
is governed by the factor of $\exp [-i \Delta t]$. Integrating
Eq. (\ref{bewgl1}) by neglecting any other time dependence of the
r.h.s. then yields an algebraical expression for $c_e(\vec{q})$
in terms of $c_\pm (\vec{q})$ which can be used to eliminate it
in Eqs. (\ref{bewgl2}) and (\ref{bewgl3}). The resulting
differential equations are
\begin{eqnarray}
  i\dot{\tilde{c}}_+(\vec{q}) &=& \frac{\Omega_{eff}}{2}
  \tilde{c}_-(\vec{q}+2\hbar \vec{k})
  e^{-i (\vec{k}\cdot \vec{a}t^2 +2 \vec{q}\cdot \vec{k}t/M
  + 4 \delta_R t)}
  +\frac{\Omega_B}{2} e^{i \Delta\varphi}
  \tilde{c}_-(\vec{q}) \nonumber\\
  i\dot{\tilde{c}}_-(\vec{q}) &=& \frac{\Omega_{eff}}{2}
  \tilde{c}_+(\vec{q}-2\hbar \vec{k})
  e^{i (\vec{k}\cdot \vec{a}t^2 +2\vec{q}\cdot \vec{k}t/M -4
  \delta_R t)} +\frac{\Omega_B}{2} e^{-i \Delta\varphi}
  \tilde{c}_+(\vec{q})
\label{2niv}\end{eqnarray}
with $\tilde{c}_\pm := \exp [i t\Omega_{eff}/2] c_\pm$. The
quantity $\Omega_{eff} := \Omega^2 /(2 \Delta)$ denotes as usual
the effective Rabi frequency of the Raman transition caused by
the lasers.

The algebraical expression for $c_e$ which we have not written down
above also
shows that $c_e$ is suppressed by a factor of $\Omega /\Delta$ as
compared to $c_\pm$. This implies that because of the very small
amount of excited atoms spontaneous emission from $|e\rangle$ to
$|\pm \rangle$ can be neglected for not too long interaction times.
To check how long the neglection of spontaneous emission is
possible we have numerically calculated the eigenvalues of the
Liouville operator ${\cal L}$ defined by
\begin{equation}
i \hbar \partial_t \rho = [\hat{H},\rho] + \Gamma \rho =:
  {\cal L} \rho
\end{equation}
where $\rho$ is the atomic density matrix, $\hat{H}$ is given
by Eq. (\ref{ham1}) if the center-of-mass degrees of freedom are
completely removed by setting $\vec{x}=\vec{p}=0$, and
the operator $\Gamma$ acts on $\rho$ by
\begin{equation}
\Gamma\rho := i\hbar \gamma \rho_{ee}(-2 |e\rangle\langle e|\,
  +\, |+\rangle\langle +| \, +\, |-\rangle\langle -|) -i \hbar
  \gamma (\rho_{e+}|e\rangle\langle +| + \rho_{e-}
 |e\rangle\langle -| + H.c. )
\end{equation}
where $2\gamma$ is the decay rate. Adopting the experimental
data of Ref. \cite{kasevich92} ($\Omega = \gamma = 10^7$ Hz,
$\Delta = 2.5\cdot 10^9$ Hz) and by setting $\varphi_\pm =
\varphi_B=0$
and $\Omega_B = 10^5$ Hz we have found that the coherences
between $|+\rangle$ and $ |-\rangle$ have a lifetime of about
1/80 second. This will turn out to be enough for our purposes.

We turn back to the coupled differential equations (\ref{2niv}).
One can read off directly that only states with the same momentum
or states which in momentum space are separated by $\pm 2 \hbar
\vec{k}$ are coupled. The first coupling is proportional to $\Omega
_B$ and is not related to a momentum transfer. The second is
proportional to $\Omega_{eff}$ and is related to a momentum
transfer $\pm 2\hbar \vec{k}$, respectively.
Because of this, Eqs. (\ref{2niv})
couple in fact a total ladder of states separated by multiples of
$2\hbar \vec{k}$ in momentum space, see Fig. 2. Because only
differences are fixed, it is evident that there is not only one
ladder but a total family of ladders. We will parametrize
different ladders by
an arbitrary momentum parameter $\vec{q}_0$ that later will denote
the measured momentum at $t=0$.

To make this structure more transparent,
and especially to clarify the
physical nature of the transition between states $|+ \rangle$ and
$|-\rangle$  of different momentum, we change from $\tilde{c}_\pm$
to the new coefficients $u_\pm$ by
\begin{eqnarray}
  \tilde{c}_+(\vec{q}_0 + 2n\hbar \vec{k}) &=:& e^{2iD_0 n t}
  e^{4i \delta_R t n^2} e^{i(n+1) \Delta\varphi}
  e^{in\vec{k}\cdot\vec{a}t^2} u_+(n) \nonumber\\
  \tilde{c}_-(\vec{q}_0 + 2n\hbar \vec{k}) &=:& e^{2iD_0 n t}
  e^{4i \delta_R t n^2} e^{in \Delta\varphi}
  e^{i n\vec{k}\cdot\vec{a}t^2} u_-(n)
\end{eqnarray}
$D_0 := \vec{q}_0\cdot \vec{k} /M$ denotes the Doppler
shift associated with $\vec{q}_0$ and + and - refer to the internal
states $|+ \rangle$ and $|-\rangle$.
$n=0,\pm 1,\pm 2,\ldots$ indicates for a state the difference
$2 n \hbar \vec{k}$ in momentum parameter relative to $\vec{q}_0$.
This leads to the final form of the dynamical equation
\begin{equation}
i \partial_t \left ( \begin{array}{c} \vdots \\
  u_+(n) \\ u_-(n) \\ u_+(n-1) \\u_-(n-1) \\ \vdots \end{array}
  \right ) =
  \left ( \begin{array}{cccccc} \ddots & \ddots & & & & \\
  \ddots & d_{n}(t) & \Omega_B /2 & & & \\ & \Omega_B /2 &
  d_{n}(t) & \Omega_{eff}/2 & & \\ & & \Omega_{eff}/2 &
  d_{n-1}(t) & \Omega_B/2 & \\ & & & \Omega_B/2 & d_{n-1}(t) &
  \ddots \\ & & & & \ddots & \ddots \end{array} \right )
  \left ( \begin{array}{c} \vdots \\
  u_+(n) \\ u_-(n) \\ u_+(n-1) \\u_-(n-1) \\ \vdots \end{array}
  \right )
\label{schroed}\end{equation}
with
\begin{eqnarray}
  d_n(t) &:=& 2 n \{ D_0 + 2 n\delta_R + \vec{k}
  \cdot \vec{a} t\} \label{dn} \\ &=&
  \frac{(\vec{q}_0 + 2n \hbar \vec{k} + M \vec{a} t)^2}{2M \hbar}
  - \frac{(\vec{q}_0 + M \vec{a} t)^2}{2M \hbar}
  \; . \label{dn2} \end{eqnarray}

The difference between the diagonal elements $d_n(t)$ and $d_{n-1}
(t)$
\begin{equation}
d_n(t) - d_{n-1}(t) = 2 [ D_0 + 2 \delta_R (2n-1) + \vec{k}\cdot
  \vec{a} t ] \label{diff} \end{equation}
varies linearly with $t$ with a factor independent of $n$.
It is at this point where the influence of the
homogeneous gravitational field appears.
Several overlapping $2\times 2$ matrices can be recognized on the
r.h.s. of Eq. (\ref{schroed}).
Based on this we can read off from Eq. (\ref{schroed}) the
following types of transitions: If the lasers are switched off
there are no Raman transitions ($\Omega_{eff}=0$). It remains the
Rabi flopping between the states $|+ \rangle$ and $|- \rangle$
caused by the magnetic field ($\Omega_B \neq 0$). There is no
accompanying
change of momentum and accordingly no coupling to states with
different $n$. If on the other hand
the magnetic field is switched off ($\Omega_B=0$)
there are transitions between $|+ \rangle$ and $|- \rangle$ with
momentum transfer $\pm 2\hbar \vec{k}$. It is essential for the
following to note that these transitions are, because of the
time dependence in Eq. (\ref{diff}), of the {\em Landau-Zener type}
if gravitation is present ($\vec{a} \neq 0$).
We will come back to this below.
If in addition $\vec{a}$ is switched off we have
the Raman transitions between $|+ \rangle$ and $|- \rangle$
with momentum transfer leading to a sort of vibration of the
center-of-mass motion. Leaving apart for the moment
the effectivity of
transitions and the question when transitions take place we obtain
for nonvanishing $\vec{a}$, $\Omega_{eff}$, and $\Omega_B$
a ladder of combined influences a section
of which is depicted schematically in Fig. 2. Let us now turn
to the details of the processes involved.
\section{Landau-Zener transitions}
A Landau-Zener transition between two states happens if in the
$2\times 2$ matrix of the responsible Hamiltonian the difference
between the diagonal terms changes linearly in time while the
non-diagonal terms remain constant \cite{landau32,zener32}.
It has been discussed for a harmonically oscillating two-level system
in \cite{garraway92}. Above in Eq. (\ref{schroed}) it is a transition
between the states $|+ \rangle$ and $| - \rangle$, below in Eq.
(\ref{dglw}) it will be a transition between dressed states. In
both cases the non-diagonal elements are proportional to $\Omega
_{eff}$ and the difference between the diagonal elements is of the
form $2 \vec{k}\cdot \vec{a} \Delta t$ with appropriate $\Delta t$.
Because of its ingredients the L.Z. transitions represent in our
case a typical {\em gravito-optical effect}. The dynamics of the
L.Z.  transition can be solved analytically. It results in a certain
efficiency of a population inversion. In our case it is
the following: If the atom is prepared in an initial state $u_-$
or $u_+$ the probability to find it in this state after the
transition has occurred is
\begin{equation}
P_{\mbox{{\footnotesize stay}}} = \exp \Big \{ -\pi
   \frac{\Omega_{eff}^2}{16 |\vec{k}\cdot \vec{a}|} \Big \}
   \; . \label{lzerg} \end{equation}
This fixes for a cloud of atoms the redistribution of the
population on the two levels. It
is an exact result which cannot be derived within perturbation
theory because of its nonanalytical dependence on
$\vec{k}\cdot \vec{a}$. In addition, it is known for L.Z.
transitions that they essentially happen at the time
when the difference of
the diagonal elements vanishes. This is in our case according
to Eq. (\ref{diff}) at the times
\begin{eqnarray}
t_n &=& - \frac{D_0 + 2\delta_R (2 n-1)}{\vec{k}\cdot \vec{a}}
   \nonumber \\
    &=& \frac{\vec{k}\cdot[\vec{q}_0 + \hbar \vec{k}(2n-1)]}{
    (-\vec{k}\cdot \vec{a})} \; .
\label{tn} \end{eqnarray}
Two subsequent L.Z. transition can happen after a time difference
\begin{equation} \Delta t :=
   t_{n+1} - t_{n} = - \frac{4 \delta_R}{\vec{k}
   \cdot \vec{a}} = - \mbox{sign}(\vec{k}\cdot\vec{a}) \frac{2
   \hbar |\vec{k}|}{M|\vec{a}|}\; . \label{deltat} \end{equation}
This is just the time during which the atom's momentum is changed
by the amount $|2 \hbar \vec{k}|$ through the influence of the
Earth's acceleration. The order of magnitude of the effective
duration of the inversion process is
\begin{equation} t_{L.Z.} := \frac{\Omega_{eff}^2}{|\vec{k}\cdot
  \vec{a}|^{3/2}}\; . \label{tlz} \end{equation}

We are now able to answer the following question: For a given initial
situation, at what time and between which states do L.Z. transitions
happen? In a first step we analyze the equations obtained and add
then below a more intuitive picture so that the underlying physics
becomes even more transparent. For simplicity we use vertically
oriented lasers: $-\vec{k}\cdot \vec{a} = k a >0$ with $\vec{a} =
-a \vec{e}_z$ whereby $\vec{e}_z$ is pointing upwards. Let us assume
that at $t=0$ the atom is prepared with initial momentum $\vec{q}_0$
in one of the states $|+ \rangle$ and $| -\rangle$ corresponding to
$u_+(n=0)$ or $u_-(n=0)$, respectively. The magnetic field will then
immediately induce Rabi oscillations between these states so that
both of them can be the starting point for a L.Z. transition,
compare Fig. 2. Possible candidates are $n=0 \rightarrow n=-1$ at the
time $t_0$ and $n=0 \rightarrow n=1$ at the time $t_1$. In any case,
a necessary condition is that the transition time $t_n$ is positive.

We distinguish three different domains for the initial momentum: (i)
For $\vec{k}\cdot \vec{q}_0 > \hbar \vec{k}^2$ Eq. (\ref{tn}) shows
that at the time $t_0$ a L.Z. transition is possible. Fig. 2
demonstrates in detail that it must be the transition $u_-(n=0)
\rightarrow u_+(n=-1)$.
From states with $n=-1$ L.Z. transitions seem to be possible to
$u_-(n=0)$ and $u_+(n=-2)$. But the corresponding times are
$t_0$ and $t_{-1}$ which are both not later than $t_0$.
Accordingly the first transition at $t_0$ remains the only one.
Note that because of the
free fall we have for the measured momentum
$\vec{p} = \vec{q}_0-M \vec{a} t$. Accordingly, $t_0$ turns out
to be  the time after which gravitation has decelerated the atom so
that the z-component of its momentum is $+\hbar k$. After
the L.Z. transition and the corresponding momentum transfer $-2
\hbar k$ it is $-\hbar k$ and the atom carries on falling without
further L.Z. transitions. This happens, too, in our next case: (ii)
$\vec{k}\cdot \vec{q}_0 < -\hbar \vec{k}^2$. From an
analysis based on Eq. (\ref{tn}) it follows that there is only the
free fall without any L.Z. transition. (iii) Finally, for
$-\hbar k^2 < \vec{k}\cdot \vec{q}_0 < \hbar k^2$, Eq. (\ref{tn})
shows that a first L.Z. transition happens for $t=t_1$. Because of
$\Delta n = +1$ we read off from Fig. 2 that it must be $u_+(n=0)
\rightarrow u_-(n=1)$. Analysing again Eq. (\ref{tn}) we see that
in this case successively at $t_2>t_1$ and then at $t_3>t_2$ and
so on L.Z. transitions from $| + \rangle$ to $| - \rangle$ take
place. Referring to the z-component of the measured momentum and
taking the free fall into account it is easy to see that $t_1$ is
the time at which the z-component of the measured momentum reaches
$-\hbar k$. The L.Z. transition transfers to a
momentum value $+\hbar k$,
the atom starts falling again until $-\hbar k$ is reached and so on.

For an intuitive understanding
it may be helpful to illustrate the analytical results derived above
in a diagram. Let us restrict for simplicity to a one-dimensional
problem in assuming the vertical
initial momentum $\vec{q}_0 = q_0 \vec{e}_z$
at $t=0$. The momentum parameter is $\vec{q} = q \vec{e}_z$.
Then the measured
momentum $\vec{p} = p \vec{e}_z$ is because of the free fall
$p= q - M a t$. The total energy
$E_{tot}$ of the atom
is the sum of the kinetic energy and the internal energy $E_\pm$
of the respective state.
$E_{tot}$ is shown in Fig. 3 as function of the measured momentum
$p$. The rapid Rabi oscillations
caused by the magnetic field are vertical
transitions between the two curves (no momentum transfer, the
dashed arrows are examples). A L.Z. transition is a transition
from the $|+\rangle$ curve to the $|-\rangle$ curve with momentum
transfer $+2\hbar \vec{k}$ or from the $|-\rangle$ curve to the
$|+\rangle$ curve with transfer $-2\hbar \vec{k}$
(solid diagonal arrows).
We know from the calculation that for the
accelerated atom the transitions caused by the
combined influence of the gravitational field and the
lasers are of the L.Z. type. They are transitions from the
$|+\rangle$ curve to the $|-\rangle$ curve with momentum
transfer $+2\hbar \vec{k}$ or from the $|-\rangle$ curve to the
$|+\rangle$ curve with transfer $-2\hbar \vec{k}$.
But for such a transition to be possible we have in addition to
fulfill the energy condition
\begin{equation} E^{tot}_+ + \hbar \omega_+ -\hbar \omega_- =
  E^{tot}_- \end{equation}
or with the specifications assumed in Eqs. (\ref{freqbed}) and
(\ref{obres})
\begin{equation} E^{tot}_+ - E^{tot}_- = E_+ - E_-
\; . \end{equation}
It can directly be read off from Fig. 3 that there are only two
transitions which fulfill the conditions of energy transfer and
momentum transfer simultaneously. They are indicated by the solid
line arrows.

We now discuss different initial conditions with the help of Fig. 3.
Let us start at $t=0$ with an initial momentum $q_0$ between $-
\hbar k$ and $+\hbar k$. We know that the Rabi transitions
caused by the magnetic field are vertical transitions between
the $|+ \rangle$ and the $|- \rangle$ curve which are permanently
happening (dashed arrows). Superimposed to this is the only slightly
disturbed free fall $p = q_0 -M a t$. Accordingly, we
move with the dashed arrows to the left in the diagram until
$p = -\hbar k$ is reached. Then the transition
from $|+ \rangle$ to $|-\rangle$ and to $p =
+\hbar k$ is in resonance and can happen (solid line arrows). The
respective probability is the one of a L.Z. transition given by
$1-P_{\mbox{{\footnotesize stay}}} $ of Eq. (\ref{lzerg}). The
influence of gravity continues. The atoms which have not made the
L.Z. transition continue to move to the left in Fig. 3 without
having another possibility for a L.Z. transition. The atoms which
have made the L.Z. transition fall freely starting with $
p = +\hbar k$ until $p=
-\hbar k$ is reached again so that a second L.Z. transition can
happen and a new cycle begins. In a similar way one can read off
from Fig. 3 that there can be only one L.Z. transition (no
cyclic behaviour) if one starts with $|- \rangle$ and
$p > \hbar k$ and that there will be
no L.Z. transition at all if one starts with $p < -\hbar k$.

It is important to check that the time (\ref{deltat}) between
two subsequent Landau-Zener transition is long enough so that each
transition can be completed. This means that the time difference
(\ref{deltat}) must be large compared to $t_{L.Z.}$. To check
whether this condition is fulfilled for optical transitions we
adopt $\Omega = 10^7 $ Hz and $ \Delta = 2.5\cdot 10^9$ Hz
from Ref. \cite{kasevich92}. The momentum $\vec{q}_0$ should be
of the order of $\hbar \vec{k}$ so that $D_0$ is of the order of
$\delta_R$. A typical value for $|\vec{k}|$ is $10^7$ m$^{-1}$ so
that for light atoms ($M \approx 10^{-26}$ kg) $\delta_R$ and
$D_0$ are of the order of $10^6$ Hz. For $\Omega_B$ we will assume
that it can be made large enough to be one order above the
effective Rabi frequency $\Omega_{eff} \approx 2 \cdot 10^4$
Hz. The last time scale is introduced by $\sqrt{|\vec{k}\cdot
\vec{a}|}$ which is of the order of $10^4$ Hz. Inserting this into
Eqs. (\ref{tlz}) and (\ref{deltat}) we see that the time between
two transition is about $0.01$ seconds whereas $t_{L.Z.}$ is of
the order of $10^{-4}$ seconds so that the transitions have enough
time to be completed.
\section{Complete solution in terms of dressed states}
Especially the discussion of Fig. 3 given above revealed that the
transitions due to the magnetic field are permanently present and
that they are necessary to enable a chain of L.Z. transitions.
It is therefore reasonable to
switch to a description of the process which incorporates these
facts from the beginning. This can be done in referring not to
$|\pm \rangle$ but to dressed states of the atom with respect to the
magnetic field. They are eigenstates of the atomic energy and the
interaction energy of the magnetic field taken together. In this
case we expect pure L.Z. transitions only.

To reduce Eq. (\ref{schroed}) to a Landau-Zener problem we
introduce the functions
\begin{equation}
w_\pm (n) := \frac{1}{\sqrt{2}} (u_+(n) \pm u_-(n)) e^{\pm i
  \Omega_B t/2}
\end{equation}
$w_\pm (n)$ correspond to wave vectors $|\psi_2
\rangle$ after the second unitary transformation of the form
\begin{equation}
|\psi_2 \rangle = | \vec{q}_0 + 2n \hbar \vec{k} \rangle \otimes
  \frac{1}{\sqrt{2}} \Big \{ e^{i \Delta \varphi} |+ \rangle \pm
  |- \rangle \Big \} e^{i\Psi (t)}
\label{wzust} \end{equation}
where $\Psi (t)$ is a time dependent phase. It is easy to show that
these wave vectors are eigenvectors of the Hamiltonian (\ref{h2})
if $\Omega =0$. Hence they are {\em dressed states} in the sense that
the effect of the (classical) magnetic field is already taken into
account.

For $w_\pm (n)$ the Schr\"odinger equation (\ref{schroed}) becomes
\begin{eqnarray}
i \dot{w}_+(n) &=& d_n w_+(n) + \frac{\Omega_{eff}}{4} \Big \{
  w_+(n+1)+w_+(n-1) + e^{i \Omega_B t} \Big [ w_-(n-1)-w_-(n+1)
  \Big ] \Big \} \nonumber \\
i \dot{w}_-(n) &=& d_n w_-(n) - \frac{\Omega_{eff}}{4} \Big \{
  w_-(n+1)+w_-(n-1) + e^{-i \Omega_B t} \Big [ w_+(n-1)-w_+(n+1)
  \Big ] \Big \} \; .
\label{dglw} \end{eqnarray}
Assuming that $\Omega_B$ is large compared to $\Omega_{eff}$ we
see that the terms proportional to $\exp (\pm i \Omega_B t)$
oscillate rapidly compared to the other coupling terms. Hence we
can perform a second rotating wave approximation in neglecting
these rapidly oscillating terms. This step decouples the $w_+$
functions from the $w_-$ functions. The new evolution equation for
the $w_+(n)$ can be written in the suggestive way as follws
\begin{equation} {\scriptstyle
i \partial_t \left ( \begin{array}{c} \vdots \\
  w_+(n+1) \\ w_+(n) \\ w_+(n-1) \\ \vdots \end{array}
  \right ) =
  \left ( \begin{array}{ccccc} \ddots & \ddots & & & \\
  \ddots & d_{n}(t)+ 2 \vec{k}\cdot \vec{a} (t-t_{n+1}) & \Omega
  _{eff} /4 & & \\ & \Omega_{eff} /4 &
  d_{n}(t) & \Omega_{eff}/4 & \\ & & \Omega_{eff}/4 &
  d_{n}(t)-2 \vec{k}\cdot \vec{a} (t-t_n)& \ddots \\
  & & & \ddots  &
  \ddots \end{array} \right )
  \left ( \begin{array}{c} \vdots \\
  w_+(n+1) \\ w_+(n) \\ w_+(n-1) \\ \vdots \end{array}
  \right )} \end{equation}
Here one can see that at the time $t\approx t_n$ the system
$w_+(n)$ and $w_+(n-1)$ indeed performs a L.Z. transition. The
coupling to other states is negligible because of the smallness
of $\Omega_{eff}/ (\vec{k}\cdot \vec{a} \Delta t)$.

Remembering that $P_{\mbox{{\footnotesize stay}}}$ gives the
probability that no transition occurs
one sees that the transition probability can be very close to one
if the exponent is of the order of ten or larger. Since
$\Omega_{eff}^2$ grows like $\Omega^4$ the laser power needs not to
be much larger than in Ref. \cite{kasevich92} to achieve this goal.
Given that this is the case it is clear that the new states
$\tilde{w}_+$ (or $\tilde{w}_-$) perform (for appropriate initial
conditions) exactly that sequence of
Landau-Zener transitions which we heuristically described in the
last section. The introduction of the dressed states made us
getting rid of the $\Omega_B$ coupling in Eq. (\ref{schroed})
that could disturb the effectiveness of the Landau-Zener
transitions.
\section{A trap for atoms with drops falling out}
To elucidate the practical significance of the considerations above
let us refer to a cloud of atoms and
consider the following initial condition: At $t=0$ atoms are
prepared with a definite momentum $\vec{q}_0$ with $|\vec{q}_0 |
<\hbar k$ in the dressed state corresponding to $w_+(n=0)$. Around
$t_1$ the first L.Z. transition happens and with the probability
\begin{equation} q := 1 - \exp \Big \{ -2\pi \Big |
  \frac{\Omega_{eff}^2}{32 \vec{k}\cdot \vec{a}} \Big | \Big \}\; ,
\end{equation}
the atoms are kicked upwards with momentum $\hbar \vec{k}$. In other
words, the fraction $(1-q)$ keeps falling and forms
the first drop while
the fraction $q$ is returned. After this all atoms are accelerated by
gravity until at the time $t_2$ atoms which have been sent upwards
at $t=t_1$ fall down with momentum $-\hbar \vec{k}$. The next
L.Z. transition may happen and the fraction $(1-q)$ of these atoms
keeps on falling forming the second drop while a fraction $q$
is kicked upwards, and so on. Accordingly we have obtained a {\em
trap} with atoms moving upwards at $t_n$ with momentum $\hbar \vec{k}
$, being decelerated by gravity to $-\hbar \vec{k}$ at $t_{n+1}$
when a fraction $q$ is "reflected" to $+\hbar \vec{k}$ and moves
upwards again. The trap is leaking. Drops of atoms are falling out at
the times $t_n$ with vertical momentum $-\hbar \vec{k}$ (pointing
downwards). The fraction of atoms which is still within the trap
at time $t$ with $t_n < t < t_{n+1}$ is
\begin{equation} P_{\mbox{{\footnotesize trap}}} = q^n \; .
  \end{equation}
The fraction which is contained in the drop number $l$ is
\begin{equation} P_{\mbox{{\footnotesize drop}}}(l) = q^{l-1}(1-q)
 \; .  \end{equation}

With regard to experimental realizations it is interesting to note
that the time when the fraction $e^{-1}$ of the atoms is still
remaining in the trap ("lifetime") is $\Delta t /\ln (1/q)$ with
$\Delta t$ of Eq. (\ref{deltat}). It is about
36 ms for the experimental data given above. A slight increase of
the laser power to $\Omega = 1.4 \cdot 10^7 $ Hz and therefore
$\Omega_{eff} = 4 \cdot 10^4$ Hz changes the lifetime to 0.5 s,
however. An idea
about the extension in space of the cloud of trapped atoms can be
obtained in the following way: It is
sufficient to consider the position classically. An atom which
just performed a L.Z. transition moves upwards with velocity
$\hbar k /M$ until after a time $\Delta t/2$ it is stopped by
gravity and begins to fall. The height difference between the two
positions is classically given by $a \Delta t^2 /8 = (\hbar k/M)^2 /
(2 a)$. Putting in the numbers given above this height difference
turns out to be about 0.5 mm.

Above we have assumed a well defined initial momentum $\vec{q}_0$.
For different values of $\vec{q}_0$ the respective times $t_1$
(and all $t_n$) differ. Lifetime and extension of the trap are not
changed. The trapped atoms have always vertical momenta between
$+\hbar \vec{k}$ and $-\hbar \vec{k}$ and atoms leak out
continuously with $-\hbar \vec{k}$.

On the basis of this picture we can easily complete our mathematical
discussion. Taking again $\vec{q}_0$ as initial momentum and
$|w_+(n=0)|^2 =1$ at $t=0$ we find that the quantum state at later
times is decomposed with regard to the states (\ref{wzust}) with
coefficients obeying at the time $t_n < t < t_{n+1}$
\begin{equation} |w_+(i)|^2 = \left \{ \begin{array}{cl}
  (1-q)q^i & \mbox{ for } 0\leq i \leq n-1 \\
  q^n & \mbox{ for } i=n\\
  0 & \mbox{ for } i<0 \mbox{ and } i>n
  \end{array} \right . \end{equation}
Based on this the mean momentum at the time $t_n < t < t_{n+1}$
turns out to be
\begin{eqnarray} \langle \psi | \vec{p} |\psi \rangle (t_n)&=&
  M \vec{a} t_n + \vec{q}_0 + 2\hbar \vec{k} \sum_{j=0}^n j
  |w_+(j)|^2  \label{meanp} \\ &=&
  \vec{q}_0^\perp + \hbar \vec{k} \Big \{ -2 n + 1 + 2q
  \frac{1-q^n}{1-q} \Big \} \nonumber \end{eqnarray}
where $\vec{q}_0^\perp $ is that part of $\vec{q}_0$ which is
orthogonal to $\vec{k}$.
These equations show how the
undisturbed free fall is decelerated.
\section{Discussion}
We have investigated a three level $\Lambda$-system
falling in a homogeneous gravitational field. To reverse the
free fall of the atom it is exposed to
two counterpropagating laser fields with wave vector $\pm\vec{k}$,
which induce Raman
transitions, and to a magnetic hyperfine field tuned to be
in resonance with the transition between the two ground states
$| + \rangle$ and $| - \rangle$.
A full quantum treatment  of the coupled external and internal
degrees of freedom is given. For sufficiently large detuning the
excited state can be eliminated adiabatically. There remains a
dynamical development which shows a.) Rabi flopping between the
states $| + \rangle$ and $| - \rangle$ caused by the magnetic field
(no accompanying change of momentum), and b.) transitions between
these states with momentum transfer $\pm 2 \hbar \vec{k}$ happening
for certain values of the center-of-mass momentum. These transitions
turn out to be of the Landau-Zener type. We have also given
a complete description of the process with reference to dressed
states (interaction energy of the magnetic field included) which is
theoretically more elegant.
In this description it becomes even more evident that the fundamental
transitions are the Landau-Zener transitions caused by the combined
influence of the lasers and the gravitational field. They
disappear if one of these influences is switched off. Accordingly
they represent a true gravito-optical effect.

The resulting center-of-mass motion is the following:
If the atom starts with momentum $|\vec{p}| < \hbar k$
it falls until $\vec{p}=-\hbar \vec{k}$ is reached and the
transition takes place. Afterwards it has obtained $\vec{p}=+\hbar
\vec{k}$, falls again until $\vec{p}=-\hbar \vec{k}$ and so on.
Periodically the gravitationally caused fall (actio)
induces in combination with the influence of the lasers an internal
transition (reactio) with momentum transfer
which turns the atom into an upward motion.
Obviously such a process can among other applications be used to
construct a coherent atom trap.

It is instructive to interpret the result also in the light of the
{\em equivalence principle}.
This method has already been used in Refs.
\cite{marzlin94,marzlin96} to explain the gravitational influence
of atoms in an interferometer and in a running laser wave.
In our case the equivalence principle states that the atom
moving in the Earth's gravitational field is equivalent to a free
atom moving in an uniformly accelerated reference frame
to which the sources of the electromagnetic fields are fixed. In this
reference frame the potential term $M \vec{a}\cdot \vec{x}$
is absent, but the Doppler effect causes the laser frequencies
to change according to
\begin{equation}
\omega^\prime_\pm =\omega_\pm \mp \vec{k}\cdot (\vec{v}
  + \vec{a} t ) \; . \label{doppl} \end{equation}
The absolute change of the magnetic field frequency $\omega_B$
is negligible for our argument. Let the atom be initially at rest
and in resonance with the lasers. If the atom now makes a transition
from $|+ \rangle$ to $|- \rangle$ it gets the momentum $2 \hbar
\vec{k}$. Consequently, due to the Doppler effect, the atom will
be out of resonance. As can be seen in Eq. (\ref{doppl}) an
acceleration $\vec{a}$ which is anti-parallel to $\vec{k}$
will decrease the momentum until the atom is in resonance again.
This is the time when the next Landau-Zener transition can happen.
Consequently the atom is periodically tuned into and out of resonance
with the laser fields.

Let us finally ask the question, what are the physical processes
which abort the sequence of Landau-Zener transitions even for
high laser powers? The most stringent restriction comes from the
breakdown of the adiabatic elimination procedure for the excited
state. As we have seen above this will happen after 1/80 second.
But this is about the time which lies between two subsequent
transitions. Hence with contemporary experimental devices we only
can hope to see the onset of the transition sequence. Another
process which weakens the efficiency of the sequence is the
coupling between $\tilde{w}_+$ and $\tilde{w}_-$ which
is neglected here after the rotating wave approximation with
respect to $\Omega_B$ in Eq. (\ref{dglw}).
Since it is likely that an $\Omega_B$ exceeding $\Omega_{eff}$
by several powers of ten cannot be produced with the present
technology this effect can be relevant. Numerical investigations
indicate that for $\Omega_B \approx 10 \Omega_{eff}$ the population
transfer from $\tilde{w}_+$ to $\tilde{w}_-$ is in the order of
10 \%. We therefore expect that this process will interrupt the
Landau-Zener sequence after a few transitions.\\
{\em Note added:} After the submission of this paper we became
aware of a preprint \cite{saprp} reporting the measurement of
Bloch oscillations of atoms. The connection between these
oscillations and the present work will be considered elsewhere
\cite{marz96b}
\\[3mm]
{\bf Acknowledgement}: We thank S. Stenholm, G. Rempe, and S.
Kunze for fruitful discussions. K.-P. M. thanks the Deutsche
Forschungsgemeinschaft for financial support.

\newpage
{\bf {\Large Figure captions}}\\[1cm]
Figure 1: A three-level atom ($\Lambda$-system)
is simultaneously exposed to two counterpropagating running
laser beams, a magnetic hyperfine field causing transitions
between the two ground states, and the gravitational field of the
Earth.
\\[1cm]
Figure 2: Pictorial representation of the ladder of influences
in Eq. (\ref{schroed}). The Landau-Zener-like Raman transitions
(L.Z.)
are connected with a momentum transfer of $2\hbar \vec{k}$
to the atom (diagonal arrows) whereas
the magnetic hyperfine transition ($\Omega_B$) induces
an interaction between states of equal momentum (horizontal
arrows). Both transitions together couple only a discrete set
of momentum states labeled by $u_\pm (n)$.
\\[1cm]
Figure 3: Total energy $E_{tot}$ of the two ground states
$| + \rangle$ and $| - \rangle$ of a $\Lambda$-system as a function
of its momentum $\vec{p}$. The Landau-Zener transition between
$| + \rangle$ and $| - \rangle$ induced by the laser fields is
connected with an energy transfer $\pm (E_+ -E_-)$ and a momentum
transfer $\pm 2 \hbar \vec{k}$. Because of the associated change in
the kinetic energy such a transition can only be in resonance
if the atom has the momentum $\vec{p}= + \hbar \vec{k}$ or
$\vec{p}=-\hbar \vec{k}$ (solid line arrow). In addition, there are
for all values of $\vec{p}$ Rabi oscillations caused by the magnetic
field (vertical dashed arrows).

\begin{thebibliography}{99}
\bibitem{mlynek94} C. Adams, M. Sigel, and J. Mlynek,
        Phys. Rep. {\bf 240}, 143 (1994).
\bibitem{marzlin96} K.-P. Marzlin and J. Audretsch, Phys.~Rev.~A.
	{\bf 53}, 1004 (1996).
\bibitem{labo95} C. L\"ammerzahl and C.J. Bord\'e, Phys. Lett.
	A {\bf 203}, 59 (1995).
\bibitem{balykin89} V.I. Balykin and V.S. Letokhov, Appl. Phys.
	B {\bf 48}, 517 (1989).
\bibitem{aminoff93} C.G. Aminoff, A.M. Steane, P. Bouyer, P.
	Desbiolles, J. Dalibard, and C. Cohen-Tannoudji, Phys. Rev.
	Lett. {\bf 71}, 3083 (1993).
\bibitem{wallis92} H. Wallis, J. Dalibard, and C. Cohen-Tannoudji,
	Appl. Phys. B {\bf 54}, 407 (1992).
\bibitem{liston95} G.J. Liston, S.M. Tan, and D.F. Walls
	Appl. Phys. B {\bf 60}, 211 (1995).
\bibitem{newbury95} N.R. Newbury, C.J. Myatt, E.A. Cornell, and C.E.
	Wiemann, Phys. Rev. Lett. {\bf 74}, 2196 (1995).
\bibitem{pritchard93} D.E. Pritchard and W.E. Ketterle, in {\em
	Laser manipulation of atoms and ions}, edited by E. Arimondo,
	W.D. Phillips, and F. Strumia, North-Holland, Amsterdam 1992.
\bibitem{klimov95} V.V. Klimov and V.S. Letokhov, Opt.
	Commun. {\bf 121}, 130 (1995).
\bibitem{moler92} K. Moler, D.S. Weiss, M. Kasevich, and S. Chu,
        Phys. Rev. A {\bf 45}, 342 (1992).
\bibitem{kasevich92} M. Kasevich and S. Chu, Appl. Phys. B {\bf
        54}, 321 (1992).
\bibitem{landau32} L.D. Landau, Phys. Zeitschrift {\bf 2}, 46
        (1932).
\bibitem{zener32} C. Zener, Proc. R. Soc. Lond. Ser. A {\bf 137},
        696 (1932).
\bibitem{garraway92} B.M. Garraway and S. Stenholm, Phys. Rev. A
	{\bf 45}, 364 (1992).
\bibitem{marzlin94} J. Audretsch and K.-P. Marzlin,
        J.~Phys.~II (France) {\bf 4}, 2073 (1994).
\bibitem{saprp} M. Ben Dahan {\em et al.}, {\em Bloch
	oscillations of atoms in an optical potential}, to appear
	in Phys. Rev. Lett.
\bibitem{marz96b} K.-P- Marzlin and J. Audretsch, in
	preparation.
\end{thebibliography}
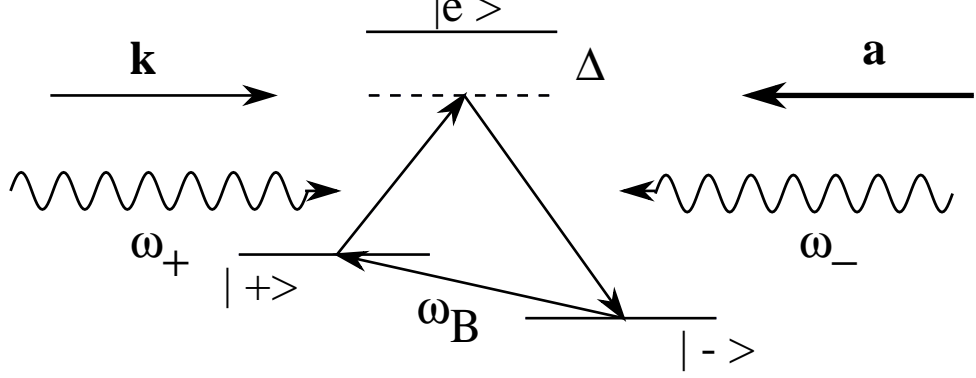
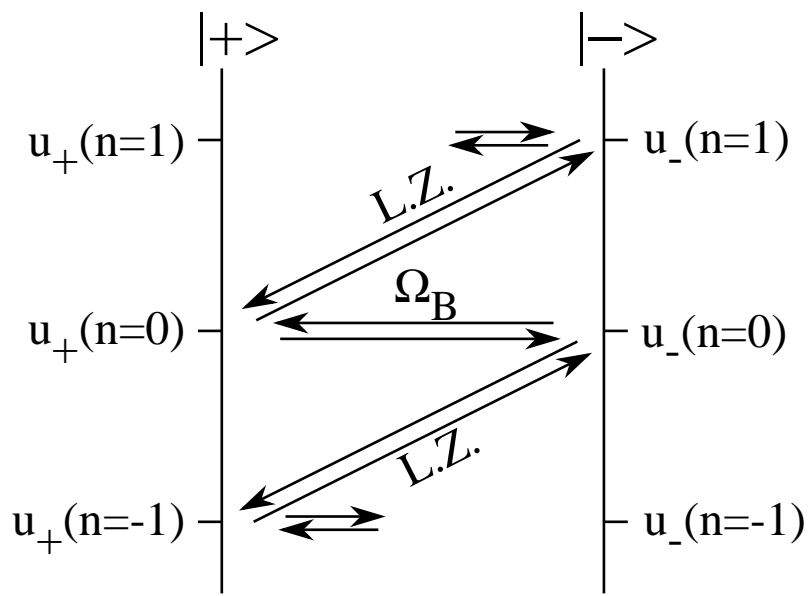
\end{document}